\documentclass[12pt,preprint]{aastex}

 \usepackage{graphicx}   

\def\ergcm2s{\ifmmode {\rm\,erg\,cm^{-2}\,s^{-1}}\else
                ${\rm\,ergs\,cm^{-2}\,s^{-1}}$\fi}

\newcommand{\lya}{\ifmmode {\rm\,Ly\alpha}\else
                Ly$\alpha$\fi}

\newcommand{\oiii}{[O\,{\sc iii}]}
\newcommand{\oii}{[O\,{\sc ii}]}

\newcommand{\lf}{erg s$^{-1}$ cm$^{-2}$}
\newcommand{\kms}{km s$^{-1}$}

\newcommand{\um}{$\mu$m}

\begin{document}
\title{\oiii\ Emission and Gas Kinematics in a Lyman-alpha Blob at z $\sim$ 3.1}
\author{Emily M. McLinden\altaffilmark{1}, Sangeeta Malhotra\altaffilmark{2},  James E. Rhoads\altaffilmark{2}, Pascale Hibon\altaffilmark{3}, Anne-Marie Weijmans\altaffilmark{4}, Vithal Tilvi\altaffilmark{5}}
\altaffiltext{1}{McDonald Observatory,  University of Texas at Austin, Austin, TX 78712, USA}
\altaffiltext{2}{School of Earth and Space Exploration,  Arizona  State University,  Tempe, AZ  85287}
\altaffiltext{3}{Gemini Observatory, La Serena, Chile}
\altaffiltext{4}{Dunlap Institute for Astronomy \& Astrophysics, University of Toronto, 50 St. George Street, Toronto, ON M5S 3H4, Canada}
\altaffiltext{5}{George P. and Cynthia Woods Mitchell Institute for Fundamental Physics and Astronomy, and Department of Physics and Astronomy, Texas A\&M University, College Station, TX, 77843}

\begin{abstract}
We present spectroscopic measurements of the \oiii\ emission line from two subregions of strong \lya\ emission in a radio-quiet Lyman-alpha blob (LAB).  The blob under study is LAB1 \citep{stei00} at z $\sim$ 3.1, and the  \oiii\ detections are from the two Lyman break galaxies embedded in the blob halo.  The \oiii\ measurements were made with LUCIFER on the 8.4m Large Binocular Telescope and NIRSPEC on 10m Keck Telescope.  Comparing the redshift of the \oiii\ measurements to \lya\ redshifts from SAURON \citep{w10} allows us to take a step towards understanding the kinematics of the gas in the blob.  Using both LUCIFER and NIRSPEC we find velocity offsets between the \oiii\ and \lya\ redshifts that are modestly negative or consistent with 0 \kms\ in both subregions studied (ranging from -72 $\pm$ 42 -- +6 $\pm$ 33 \kms). A negative offset means \lya\ is blueshifted with respect to \oiii\, a positive offset then implies \lya\ is redshifted with respect to \oiii. These results may imply that outflows are not primarily responsible for Lyman alpha escape in this LAB, since outflows are generally expected to produce a positive velocity offset \citep{mc11}. In addition, we present an \oiii\ line flux upper limit on a third region of LAB1, a region that is unassociated with any underlying galaxy.  We find that the \oiii\ upper limit from the galaxy-unassociated region of the blob is at least 1.4 -- 2.5 times fainter than the \oiii\ flux from one of the LBG-associated regions and has an \oiii\ to \lya\ ratio measured at least 1.9 -- 3.4 times smaller than the same ratio measured from one of the LBGs.
\end{abstract}


\section{INTRODUCTION}
Lyman-alpha (\lya) first became a useful tool for observing high-z sources with the discovery of large samples of \lya\ emitting galaxies (LAEs) \citep{ch98,hu98,rhoje00}.  The same narrowband imaging techniques that uncovered LAEs began uncovering a different set of objects that were also very bright in \lya.  These rarer, more extended, and more luminous objects are what we now call \lya\ blobs (LABs) \citep[e.g.,][]{stei00,fran01,mat04,dey05,nil06,pres12}.  LABs are extremely large ($\sim$ 30--200 kpc) radio-quiet \lya\ nebulae in the high redshift universe.  LABs are  highly luminous (L$_{Ly\alpha} \sim $ 10$^{43-44}$ ergs s$^{-1}$), and yet despite rigorous study in the last decade, the mechanism(s) that power this immense \lya\ flux is not fully understood.  This paper will focus on investigating the kinematics of and mechanisms powering such objects by investigating LAB1, a z $\sim$ 3.1 LAB first discovered by \citet{stei00}.

There are currently three most widely discussed scenarios to explain  both the large spatial extent and powerful \lya\ flux of these blobs.  The first of these is that the gas in LABs is heated by photoionization from massive stars and/or AGN \citep{ge09}.  A second scenario proposes that gas in LABs is heated by cooling flows / cold accretion \citep{hai00,dij09}.  Finally some authors have proposed LABs originate from overlapping supernova remnants from massive stars after a powerful starburst \citep{tan00,oh03} producing superwinds.

Adding to the controversy, observations in recent years from different authors have led to different conclusions about which of these scenarios are responsible for said observations.  \citet{nil06} have argued that their observations of a z $\sim$ 3.16 LAB were best matched by cooling flows onto a dark matter halo.  This is in contrast to the conclusions of \citet{hay11}, who found evidence of polarized \lya\ radiation in LAB1.  The \citet{hay11} results suggest that \lya\ photons are scattered at large radii from their production sites and this observation seems to not only favor the role of scattering in outflows in LABs, but the authors contend their discovery can actually rule out most inflow models.  But a similar study by \citet{pres11} found no evidence of polarization in a LAB at z $\sim$ 2.656 and these authors argue their results are inconsistent with spherical outflows and \lya\ scattering from nearby AGN.  Yet another conclusion is reached by \citet{ya11} whose observations of \lya\ and H$\alpha$ emission in two z $\sim$ 2.3 LABs rule out simple infall models and models that rely on large outflows.  This diversity of conclusions may mean that there are diverse mechanisms powering different blobs (or multiple mechanisms at play in single blobs) or it may mean that truly conclusive observations have not yet been presented.

To try to provide new data to differentiate amongst the possible LAB sources we focus on LAB1 \citep{stei00}.  LAB1 resides in SSA22, a protocluster region at z $\sim$  3.1 \citep{stei98}.  LAB1 extends $\sim$ 100 kpc \citep{w10} and has a \lya\ luminosity of 1.1 $\times$ 10$^{44}$ erg s$^{-1}$ \citep{mat04}.  This makes LAB1 one of the brightest and most spatially extended LABs yet observed \citep{stei00,mat04}.  LAB1 is comprised of five separate regions of \lya\ emission known as C11, C15, R1, R2, and R3 (see Figure 1 of Weijmans et al (2010)).  C11 and C15 are both Lyman break galaxies (LBGs) identified by \citet{stei00}.  R3 has been identified as an extremely red galaxy \citep{ge07} and may or may not be associated with a bright submillimeter source and nearby radio source (Chapman et al. 2001, Chapman et al. 2004, but see also Kohno et al. 2008, Yang et al, 2012).  R1 and R2 are not identified with galaxies \citep{w10}.   \citet{ge09} also demonstrated that AGN activity is not significant in LAB1 as LAB1 is not detected in a 400 ks Chandra exposure.  We note that \citet{mat04} identified C15 as a separate LAB, but in this paper we refer to all five subregions as components of LAB1, following the nomenclature of \citet{stei00} and \citet{w10}.

In this paper we present new spectroscopic \oiii\ observations of LAB1.  We focus on the two \lya\ subregions C15 and C11, the two parts of the larger LAB1 structure in which we detected \oiii.  The use of \oiii\ to study the kinematics of LAB1 is powerful because [O\,{\sc iii}]  is not subject to resonant scattering as \lya\ is.  Comparing \oiii\ to \lya\ emission allows us to characterize any systemic offsets between \lya\ and [O\,{\sc iii}], and thereby detect the presence of outflows or inflows.  We first demonstrated the efficacy of this method in a sample of LAEs in \citet{mc11}.

Our new \oiii\ data presented here is compared to \lya\ data originally presented in \citet{w10}.  \citet{w10} measured \lya\ from each of the 5 subregions in LAB1 with the integral field spectrograph SAURON over 23.5 hours (including 9 hours of SAURON data originally obtained by \citet{bow04}).  We obtained the reduced datacube produced from these observations from Weijmans.  This allows us to extract \lya\ profiles at locations corresponding to our NIR \oiii\ observations for careful comparison of \oiii\ and \lya\ redshifts. Henceforth we refer to \citet{w10} as W10. 

In Sections \ref{nirdata} and \ref{lucidata} we present our \oiii\ detections from two near-infrared (NIR) spectrographs (NIRSPEC and LUCIFER).  In Section \ref{results} we look for any velocity offsets between our measured \oiii\ redshifts and those of \lya\ to look for any evidence of inflows or outflows and in Section \ref{discussion} we discuss the implications of our findings of velocity offsets ($\Delta$v) that are modestly negative or consistent with zero.  We also compare our results to other authors and explore if there are currently any \lya\ radiative transfer models that can match our results.   Where relevant, we adopt the standard cosmological parameters H$_{0} =$ 70 km s$^{-1}$ Mpc$^{-1}$, $\Omega_{m} = $ 0.3, and $\Omega_{\Lambda} =$ 0.7 \citep{sper}.  We use the following vacuum wavelengths, 1215.67  \AA\ for \lya, 3727.092/3729.875 \AA\ for [O\,{\sc ii}], 4862.683 \AA\ for H$\beta$, and 4960.295/5008.240 for [O\,{\sc iii}] from the Atomic Line List v2.04\footnote{http://www.pa.uky.edu/$\sim$peter/atomic/index.html}.

\section{NIRSPEC DATA AND REDUCTION}\label{nirdata}
We initially detected \oiii\ emission from LAB1 using the near-infrared spectrograph NIRSPEC \citep{mc98} on the 10m Keck~II telescope on 6 August 2010 (UT).  We used the low-resolution mode of NIRSPEC, with the 42 $\times$ 0.76\arcsec\ slit and the NIRSPEC-6 filter which covers 1.558 -- 2.315 \um.  The spectral resolution of NIRSPEC in this setup is $\sim$ 290 \kms. This filter encompasses the redshifted (z $\sim$ 3.1) \oiii\ doublet and  covers redshifted H$\beta$ as well, though we did not see this line.  We completed three 500-second integrations, for a total exposure of 25 minutes.  The longslit was oriented so that LAB1 regions C15 and C11 \citep{stei00} both lie directly on the slit.  Region R2 also has some peripheral coverage, though not directly through the location of its peak \lya\ emission.  Due to the short length of slit we were unable to place an additional continuum-bright object on the slit, so LAB1 was acquired via blind-acquisition from a nearby star.

\oiii\ detections from C15 and C11 are evident in single, raw 500-second exposures, when a second frame is subtracted from the frame of interest. We find these detections at their expected locations in the spatial direction along the slit, and find they also have the separation from one another that we expect for C11 and C15.   In addition, the detections are in the wavelength range expected of each region's \lya\ redshift.  The dither pattern we used is also clearly visible in positive and negative detections when we perform this sort of quick-look subtraction, providing assurance that these detections are not transient cosmic rays in a single exposure.  These facts combined give us confidence that the emission we detect is, in fact, from \oiii\ emission from C15 and C11.  See the top panel of Figure \ref{fig:nirplots} for these detections in a sky-subtracted 2D frame. The \oiii\ emission from C15 is strong and comes through as such in both our 2D and 1D reduction processes detailed below.  The \oiii\ detections from C11 appear much fainter and are not as evident, though still marginally detected, throughout our 2D and 1D reduction processes.

Initial data reduction of each 500-second exposure was done using NIRSPEC\_REDUCE, a set of IDL programs written by G. D. Becker specifically for reduction of NIRSPEC longslit data.  We used these scripts for flat fielding and sky subtraction in each exposure.  The sky subtraction process in NIRSPEC\_REDUCE is based on the algorithm of D. Kelson \citep{kel03} which provides excellent sky subtraction of even tilted skylines, such as those in NIRSPEC longslit data.  The outcome of these reduction steps are three individual, 2D, flat-fielded, sky-subtracted exposures.  We also reduced an argon lamp exposure and a standard star exposure in this  same way.

We then fed these exposures into IRAF procedures in the WMKONSPEC package\footnote{http://www2.keck.hawaii.edu/inst/nirspec/wmkonspec.html}.  Each frame was rectified to a horizontal-vertical grid in x- y using the tasks XDISTCOR AND YDISTCOR, which remove x and y distortion in the images, respectively.  Once the exposures were rectified, we used IMCOMBINE to median combine the frames into a single exposure.  We specified offsets in the IMCOMBINE procedure to remove dithers along the slit that were performed during our observations.  The result, what we call our reduced 2D-spectrum, is shown in Figure \ref{fig:nirplots}.  The locations of our \oiii\ detections in C11 and C15 are circled.  

\begin{figure}
\begin{tabular}{c}
\includegraphics[scale=10]{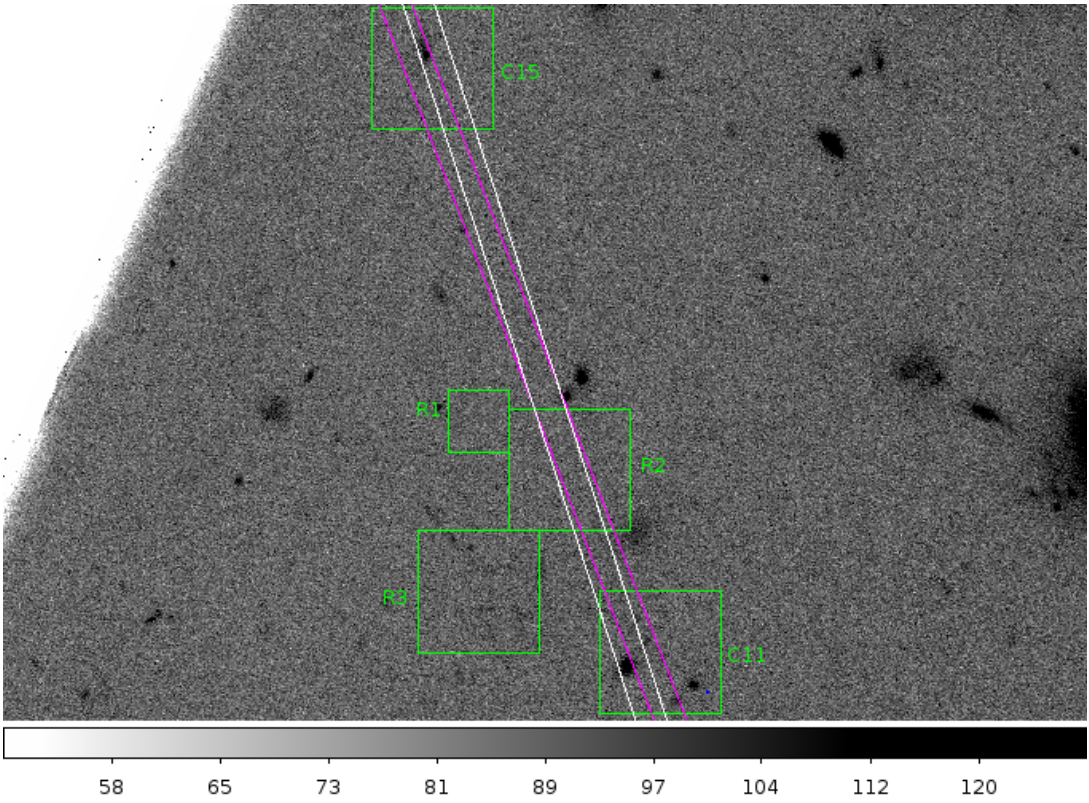}\\
\end{tabular}
\caption{Orientation of longslit for NIRSPEC and LUCIFER observations.  NIRSPEC slit is in white, LUCIFER slit is magenta.  Boxes indicate locations of subregions C11,C15, and R1-R3, from W10.  Image from STIS-MIRVIS observations, retrieved from HST archives \citep{chap03}.}\label{fig:orient}
\end{figure}

\begin{figure}
\begin{tabular}{cc}
\multicolumn{2}{c}{\includegraphics[scale=1.0]{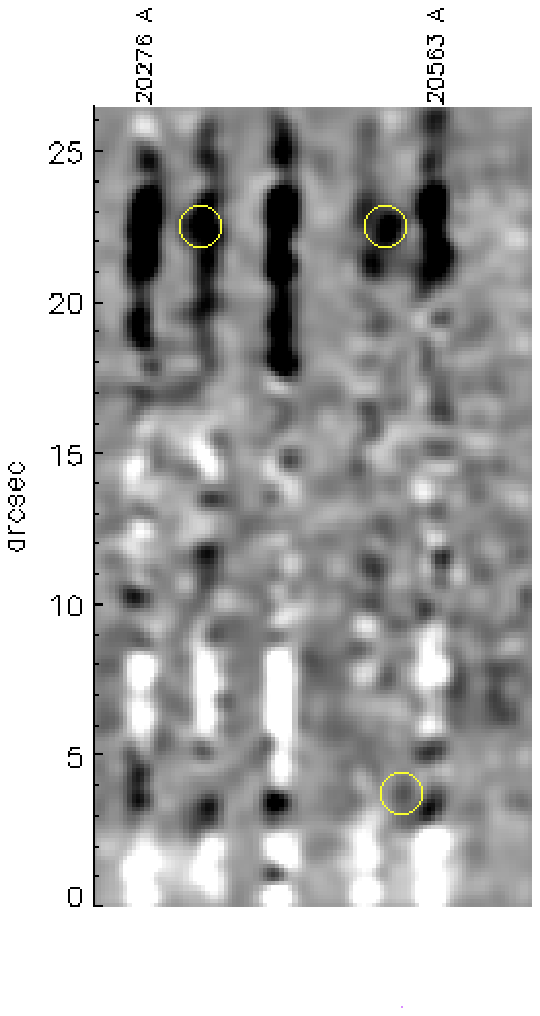} }\\
\includegraphics[bb=50 5 686 424,scale=0.35]{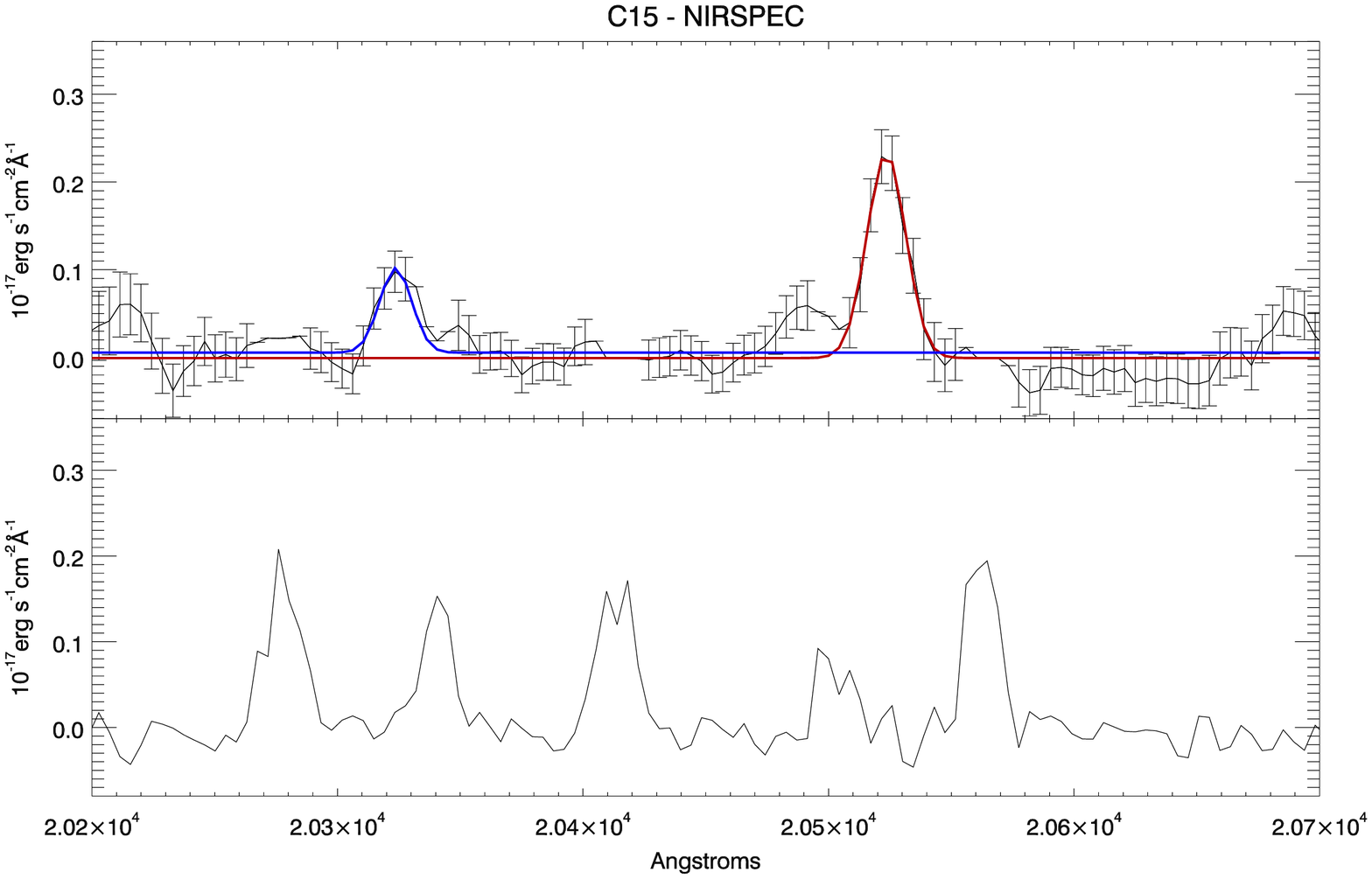} & \includegraphics[bb=90 5 686 424,scale=0.35]{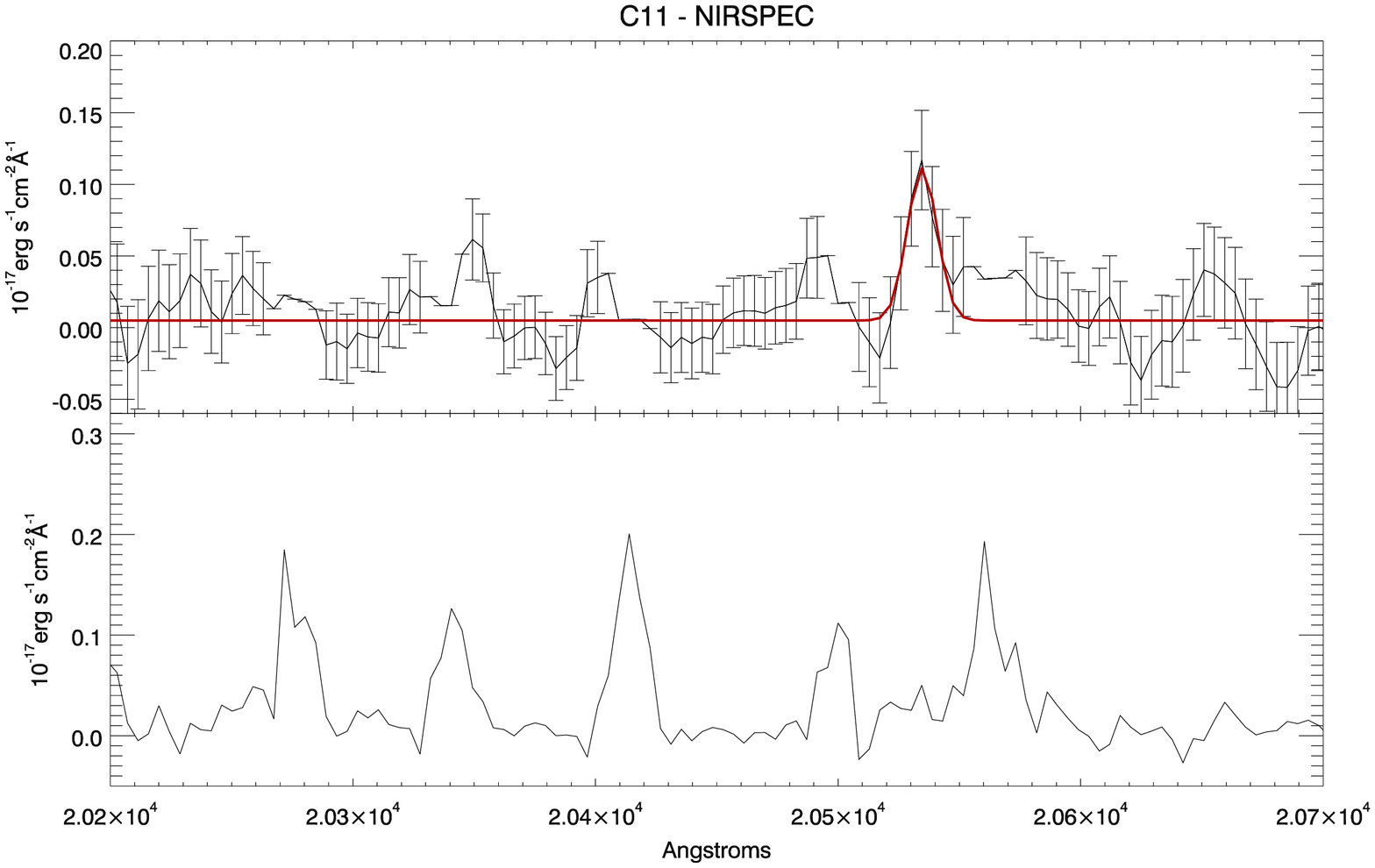}\\
\end{tabular}
\caption{Top image is median combined, sky-subtracted, distortion corrected 2D spectrum from NIRSPEC.  Top image has been smoothed with a Gaussian function with radius of three pixels.  Emission lines from C11 and C15 are circled.  \oiii\ doublet from C15 are upper circles, 5008.24 \AA\ line from C11 is lower circle.  Bottom row contains 1D NIRSPEC spectra for C15 and C11 (top panel) and extracted background.  Best-fit Gaussians are overlaid on emission lines - see Section \ref{oiiifit}. Blue is 4960.295 \AA\ line and red is 5008.24 \AA\ line.  Extracted background has been multiplied by -1 so one can easily  recognize the sky lines, but the sky lines were actually negative in some places due to imperfect sky subtraction in the 2D image.}\label{fig:nirplots}
\end{figure}

To extract 1D spectra we used the IRAF DOSLIT procedure.  We first defined an aperture trace using a bright standard star observation which can be easily traced along the entire slit.  We then transferred this aperture to the correct spatial locations in our science exposure to extract spectra of C15 and C11.  We used this transferred aperture since neither region has continuum emission that we are able to trace for aperture creation.  The spectra were wavelength calibrated using an argon lamp.  The RMS error from wavelength calibration was 0.67 and 0.45 \AA\ for C15 and C11 respectively.  We used the IRAF WMKONSPEC task SKYINTERP to remove remaining residuals from sky lines.  The resulting 1D spectra from C15 and C11 are shown in Figure \ref{fig:nirplots}.

We flux calibrate our NIRSPEC spectra using a magnitude 8.85 (V band) A1V star that was observed in the same setup as our science observations (0.76$\arcsec$ slit).  We averaged an A0V and an A2V Pickles model spectrum to approximate an A1V spectrum.  We then scaled down the A1V stellar spectrum \citep{pick} to match the magnitude of the observed star.  We created a sensitivity function with units of erg cm$^{-2}$ \AA$^{-1}$ counts$^{-1}$ by dividing the model spectrum by the observed stellar spectrum and multiplying by the length of the observation.  This sensitivity function is then multiplied by the extracted NIRSPEC spectra for C15 and C11, and then the result is divided by the integration time for each object to produce flux calibrated spectra in units of erg s$^{-1}$ cm$^-2$ \AA$^{-1}$. 

\begin{figure}[!hb]
\epsscale{0.85}
\plotone{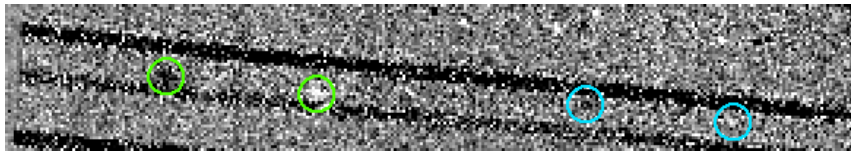}
\caption{Third NIRSPEC frame subtracted from second NIRSPEC frame, before x-axis and y-axis distortion correction, and sky subtraction.  \oiii\ emission (5008.24 \AA\ line) from C11 and C15 is more evident than in Figure \ref{fig:nirplots}. Emission from C15 in cyan circles, emission from C11 in green circles.  Emission from second frame is black (positive), emission from third frame is white (negative).  Positive-negative dither pattern is clear, showing a detection in both frames displayed here.}\label{fig:figsub}
\end{figure}

\section{LUCIFER DATA AND REDUCTION}\label{lucidata}
 We subsequently made additional NIR observations of LAB1 using LUCIFER (LBT NIR Spectrograph Utility with Camera and Integral-Field Unit for Extragalactic Research) on the 8.4m LBT \citep{sei02,ag10}.    We used the longslit mode of LUCIFER with a 1$\arcsec$ slit utilizing the H+K grating with 200 lines/mm and the N1.8 camera.   LUCIFER has a spectral resolution of $\sim$ 318 \kms\ near 1.6 $\mu$m and $\sim$ 233 \kms\ near 2.2 $\mu$m.   We combine three 300-second integrations, for a total exposure of 15 minutes.  We placed the longslit at a slightly different orientation than our NIRSPEC observations in hopes of capturing more emission from R2 and C15.  For the LUCIFER setup, C15 and R2 lay directly on the slit, with part of C11 on the edge of the slit (see Figure \ref{fig:orient}).  Given the length of the LUCIFER longslit, we were able to place a galaxy with continuum on the slit as well.  This aids in aperture extraction during the reduction process.

\begin{figure}[!hb]
\centering
\begin{tabular}{cr}
\includegraphics[bb=70 40 434 320,scale=0.5]{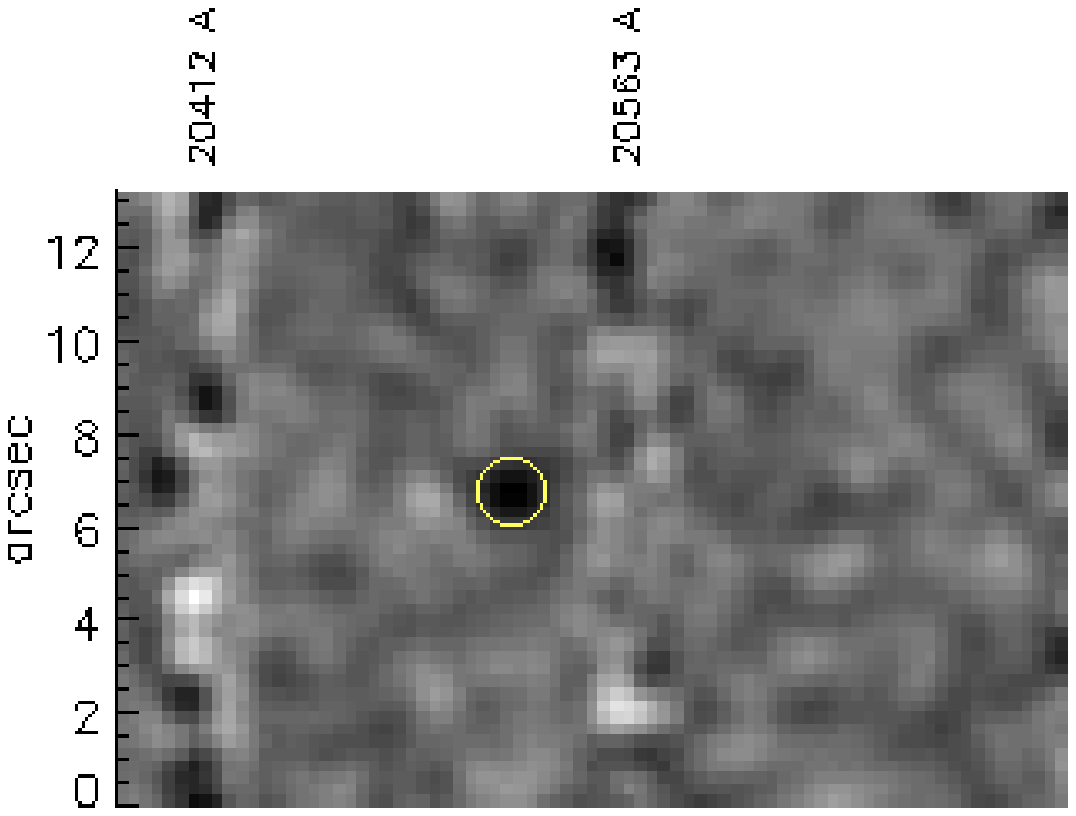} & \includegraphics[scale=0.4]{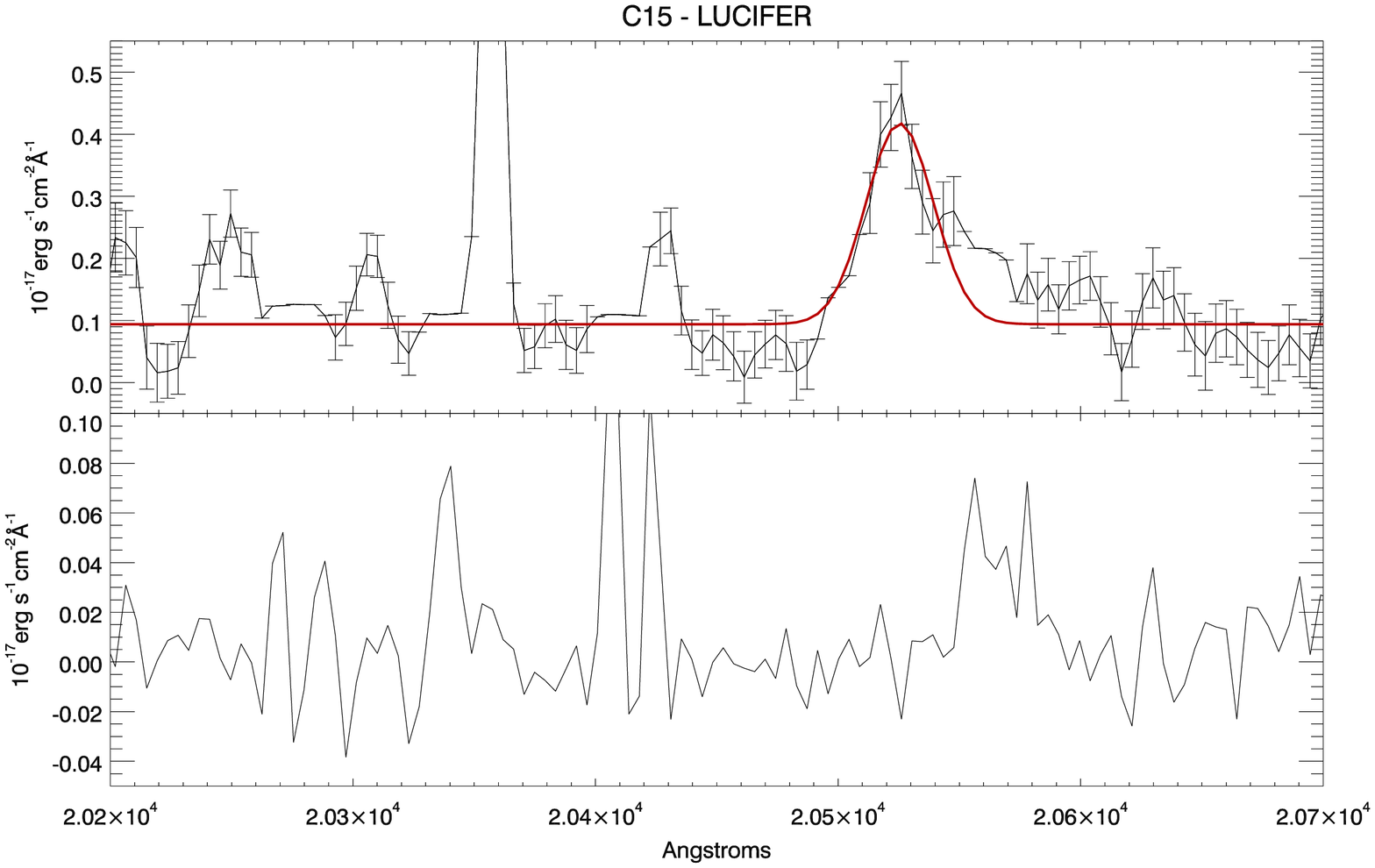}\\
\end{tabular}
\caption{Left plot is 2D LUCIFER spectrum of C15.  The image has been smoothed by a Gaussian with a radius of 3 pixels and the \oiii\ detection (5008.24 \AA) is circled.  Right top panel is extracted 1D LUCIFER spectrum of C15.  Feature at $\sim$ 20355 \AA\ is a bad column.  Best-fit Gaussian is  overlaid on emission line - see Section \ref{oiiifit}. Bottom right panel is extracted background - only minimal sky line residuals remain because sky subtraction in 2D spectrum was excellent.  Extracted background had been multiplied by -1.}\label{luciplots}
\end{figure}

We reduced the LUCIFER data in a very similar manner as the NIRSPEC data, but we used a modified version of the NIRSPEC\_REDUCE package to accommodate the different detector size and orientation of the LUCIFER data.  After the NIRSPEC\_REDUCE procedures, the individual exposures were again median combined with IMCOMBINE and offsets from dithering along the slit were accounted for.  Figure \ref{luciplots} shows the combined 2D spectrum after this step.  For the 1D extraction, we created an aperture trace using the continuum source that shared the slit with our science targets, instead of the standard star as in our NIRSPEC procedure. Then we shifted the aperture to the correct spatial locations to extract 1D spectra for C15, C11 and R2.  The spectra were again wavelength calibrated with an argon lamp exposure.  Figure \ref{luciplots} shows the 1D extraction of C15 from our LUCIFER data, which was the only region from which we detected \oiii\ in our LUCIFER observations.  We flux calibrate our LUCIFER spectra using a magnitude 6.16 (V band) A5V star that was observed in the same setup as our science observations (1$\arcsec$ slit).  We used a Pickles A5V model spectrum and otherwise the calibration process is otherwise the same as described in Section \ref{nirdata}. 


\section{\lya\ DATA}
We extracted \lya\ profiles from the reduced SAURON datacube, data that was initially presented in W10.  The SAURON data was registered to the R band image that was used to align our NIRSPEC and LUCIFER slits.  Next, we laid our LUCIFER and NIRSPEC slits  (Figure \ref{fig:orient})  on the registered SAURON data and extracted the spaxels in the SAURON data that corresponded with the LUCIFER and NIRSPEC \oiii\ detections.  This is essentially creating a virtual longslit aperture for the SAURON data.  The LUCIFER longslit was 1\arcsec\ wide and the NIRSPEC longslit was 0.76\arcsec\ wide, so our virtual longslit for the SAURON data was 3 spaxels across, or 1.2\arcsec\ to provide an approximate match to the LUCIFER and NIRSPEC apertures.   The extracted spaxels were summed to create 1D spectra for C11 and C15.  Because we observed C15 with both NIRSPEC and LUCIFER, with slightly different slit orientations, we extracted two spectra for C15 - one that corresponds to our NIRSPEC slit orientation, and one that corresponds to our LUCIFER slit orientation.  When discussing results for C15 below we specify whether the results are for C15 from the NIRSPEC alignment or C15 from the LUCIFER alignment.   C11 was only detected with NIRSPEC, so we only extracted one spectrum for C11.  The \lya\ lines detected in the 1D spectra described above are shown in Figure \ref{asymplots}.

To get error bars for the flux in the extracted 1D spectra we selected 1000 random spaxels in the SAURON datacube that were outside the area of the blob.  At each wavelength we created a histogram of the 1000 flux values found in our random spaxels at that wavelength, fit a Gaussian to that histogram, and adopted the sigma of that Gaussian as the 1 sigma error on the flux at that wavelength.

\section{RESULTS}\label{results}

\subsection{\oiii\ Redshifts}\label{oiiifit}
As we did in \citet{mc11} we fit detected \oiii\ lines with a single symmetric Gaussian plus a constant, using the IDL routine MPFITEXPR\footnote{http://cow.physics.wisc.edu/~craigm/idl/down/mpfitexpr.pro}. We fit the NIRSPEC and LUCIFER spectra independently.  The central wavelength of the best fit Gaussian determines the \oiii\ redshift.    We fit the 4960.295 \AA\ and 5008.24 \AA\ lines independently  for C15 in NIRSPEC and we find only the 5008.24 \AA\ line in the LUCIFER data for this galaxy. Given the redshift of this region from the 5008.24 \AA\ line in LUCIFER, the 4960.295 \AA\ line should fall at $\sim$ 20332.3 \AA, right on the edge of the 20339.497 \AA\ (vacuum) OH emission line \citep{rou00}.  This may explain why, after sky interpolation we are unable to detect this line in the slightly lower resolution of LUCIFER.  The agreement between the redshift derived from 4960.295 and 5008.24 \AA\ lines in NIRSPEC spectrum is good (see Table \ref{tblz}).  We take the average of the 4960.295 \AA\ and 5008.24 \AA\ redshift as the derived systemic redshift for C15 from NIRSPEC, and use the redshift of the single line for C15 from LUCIFER. 

Only the stronger 5008.24 \AA\ line was detected in C11, and only in the NIRSPEC spectrum, so the 5008.24 \AA\ line alone defined the redshift for this region.  As mentioned in Section \ref{lucidata} the location of longslit in the LUCIFER setup was optimized for detection of R2 and C15, so it is not surprising that we did not have a detection for C11 in the LUCIFER data.  All these \oiii\ redshifts were corrected for the earth's motion using topocentric radial velocities\footnote{http://fuse.pha.jhu.edu/support/tools/vlsr.html} appropriate for the date and location of the observations.   The error bars on the redshift are a compilation of the RMS from wavelength calibration during data reduction, the 1 sigma error on the best-fit central wavelength from Gaussian fitting, averaging of two redshifts when applicable, and a 0.02 \kms\ uncertainty on the topocentric radial velocities.  See Table \ref{tblz} for a summary of this data.

\subsection{\lya\ Redshifts}\label{lyafit}
\begin{figure}[!bh]
\begin{tabular}{cc}
\includegraphics[scale=0.3]{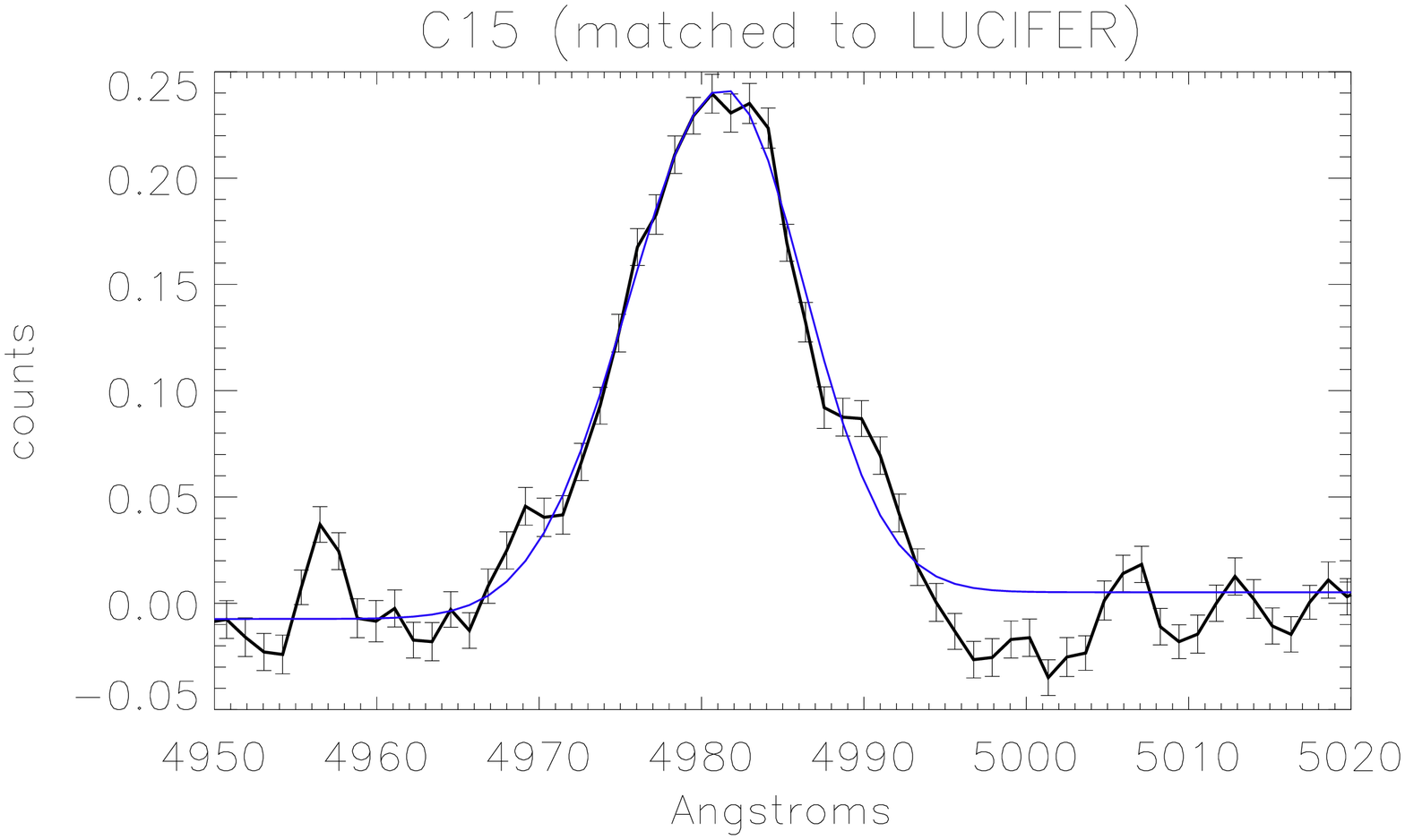} & \includegraphics[scale=0.3]{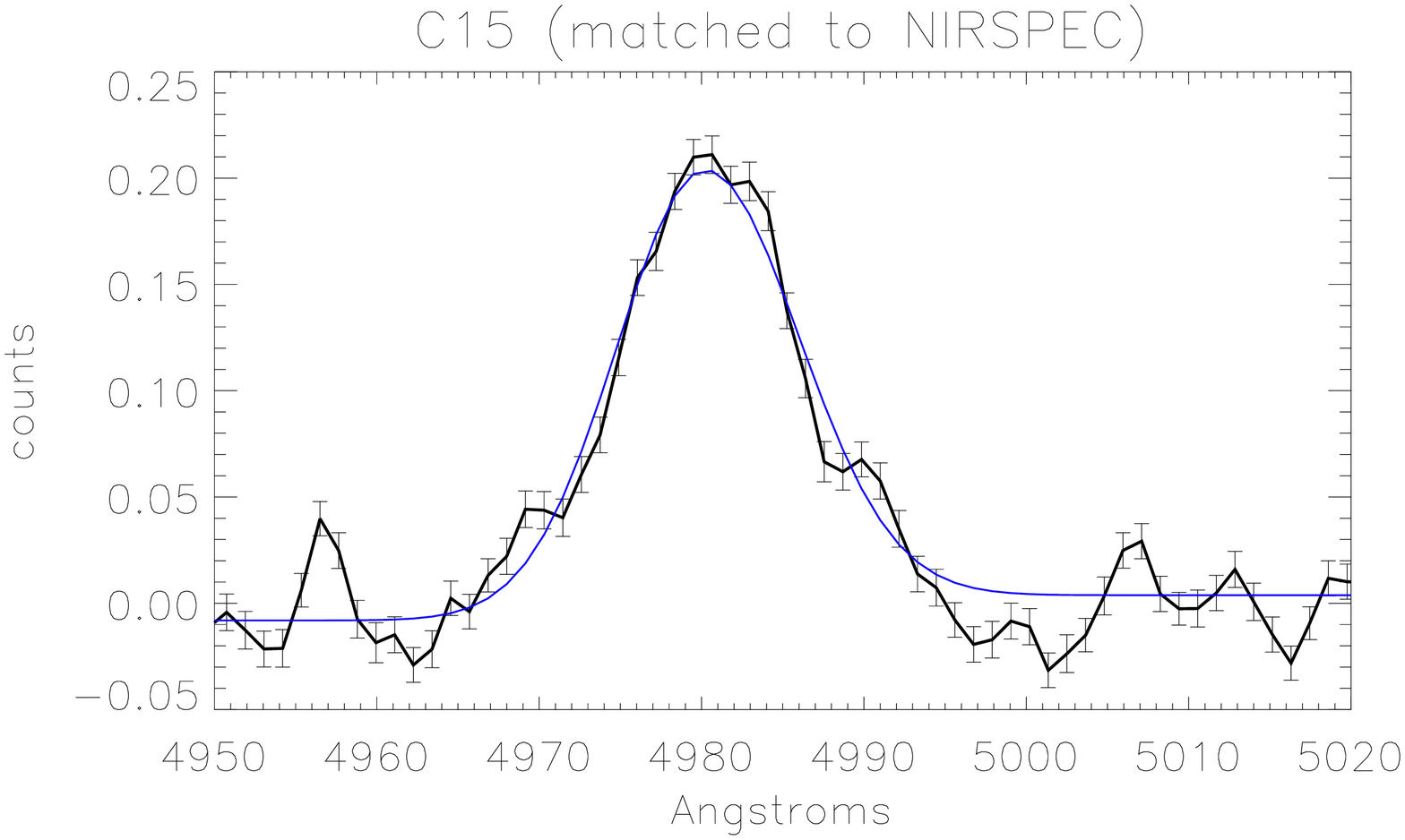}\\
\includegraphics[scale=0.3]{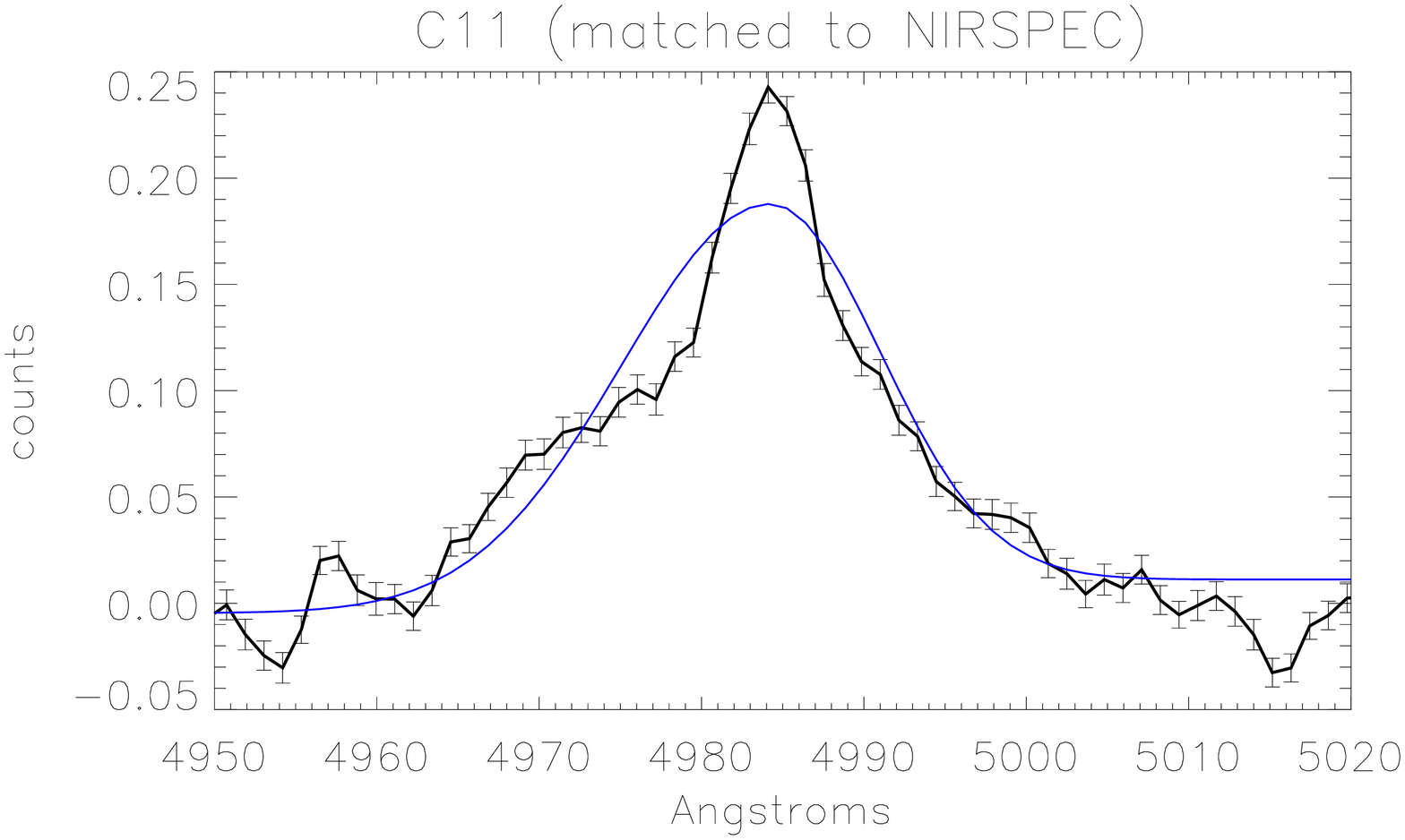} & \includegraphics[scale=0.3]{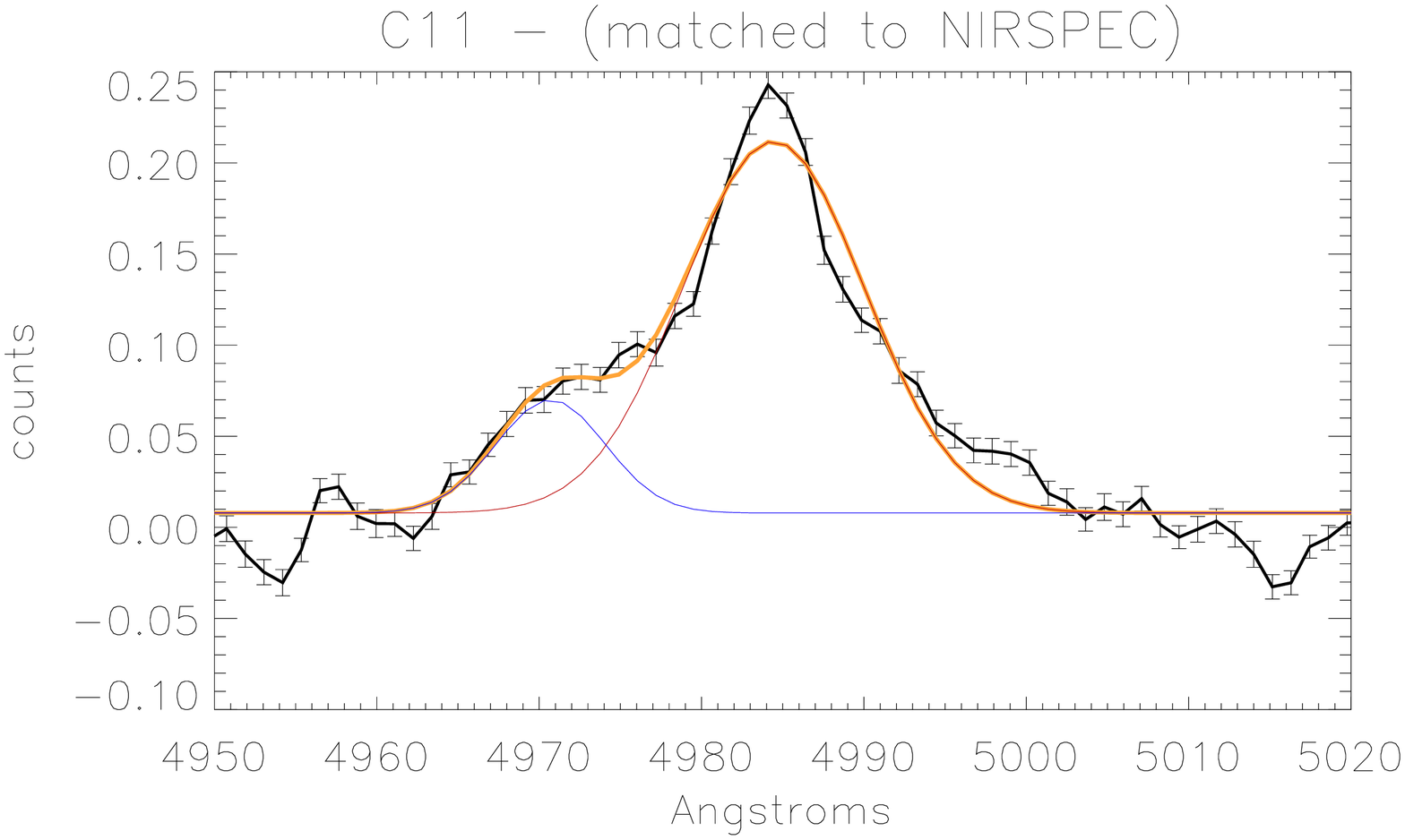} \\
\end{tabular}
\caption{Best fit Gaussians for C15 (top row) and C11 (bottom row) from ARM\_ASYMGAUSSFIT show in blue.  \lya\ spectra  are shown in black.   Bottom right panel shows C11  simultaneously fit with a double Gaussian using MPFITEXPR - the two Gaussians are shown in blue and red, and their sum is shown in orange.}\label{asymplots}
\end{figure}

To determine the redshift of the \lya\ emission line, we followed the same methodology we previously used in \citet{mc11}.  Namely, the \lya\ profiles for C11 and C15  were fit with a single asymmetric Gaussian plus constant, using IDL routine ARM\_ASYMGAUSSFIT\footnote{http://hubble.as.arizona.edu/idl/arm/arm\_asymgaussfit.pro}.  These fits are overlaid on the extracted 1D \lya\ spectra in Figure \ref{asymplots}.   The asymmetric Gaussian fitting routine allows for, but does not require, that the fit be asymmetric.  The central wavelength of the best-fit asymmetric Gaussian defines the \lya\ redshift.

The error on the line center was determined from 1000 Monte Carlo iterations.  In each iteration the flux at each wavelength was altered by a random amount proportional to the error bar at that point and the altered spectrum was re-fit with an asymmetric Gaussian.  The standard deviation of the 1000 best-fit line centers is used as the 1$\sigma$ error on the \lya\ line center.

We note that the SAURON IFU datacube comes from a combination of data obtained in 2002 \citep{bow04} and 2006 \citep{w10}.  The data were not corrected to the local standard of rest (LSR) before combination into a single datacube.  Therefore, the \lya\ profiles derived from the combined data may be slightly broadened with a slightly shifted center compared to the intrinsic \lya\ profile.  To account for this effect, we calculated the corrections that should be applied to the 2002 ($\sim$ -26.6 \kms) and the 2006 data ($\sim +$2.1 \kms)\footnote{http://fuse.pha.jhu.edu/support/tools/vlsr.html}.  Combining two such Gaussians, offset from one another by $\sim$ 28.7 \kms,  means that the intrinsic line center (if they had both been calibrated to the LSR before combination) is shifted by $\sim$ 12.3 \kms.  We therefore fold this 12.3 \kms\ term into our error calculations for redshift of the \lya\ line and the subsequently derived \oiii-\lya\ velocity offsets.   Table \ref{tblz} summarizes our \lya\ and \oiii\ redshift information.

\subsection{Velocity Offsets}\label{voff}
We derive a velocity offset ($\Delta$v) between the \oiii\ and \lya\ lines by  comparing the redshifts for each line as derived from the central wavelength of the best-fit Gaussians to the two lines, where the fitting procedure is described above.  We consider the \oiii\ redshift to be the systemic redshift.  This method finds an offset between \oiii\ and \lya\ in  C15 of -72.4 $\pm$ 41.6 \kms\ from the LUCIFER data and -51.3 $\pm$ 42.1 \kms\ from the NIRSPEC data.  These results are remarkably consistent with one another, in spite of being measured in slightly different apertures with different instruments.  These results are  modestly negative though also very nearly consistent with zero.   We find in $\Delta$v is 5.8 $\pm$ 32.9 \kms\ for C11 in the NIRSPEC data, which is consistent with no offset.  See Table \ref{tblz} for a compilation of these results.

One may argue that the rather broad \lya\ line in C11 may be better fit with a double Gaussian profile, especially if one considers that the bump to the left  of the highest peak is not noise, but in fact evidence of a second, unresolved peak.  This could be a  blue bump that is not fully resolved and separated from the main red peak.  Or it could be the smaller red peak at v=0, as in Fig. 12 in \citet{v06}.  To consider these possibilities, we fit C11 a second time, simultaneously fitting two Gaussians plus a constant (see bottom right panel of Figure \ref{asymplots}).  When we do this we find that the right peak yields only a modest offset, where the \lya\ line is offset by $\sim$ 18 \kms\ from the \oiii\ redshift of C11 from NIRSPEC.   The bluer \lya\ peak is blueshifted with respect to \oiii\ by $\sim$ 808 \kms. This leads to a rather inexplicably large offset between the two \lya\ peaks, $\sim$ 826 \kms, especially when the red peak is so mildly offset, which may disfavor this secondary interpretation of the modest bump as a blue bump.  While acknowledgement of the additional fit described above is worthwhile, the discussion stays much the same, the observed velocity offset between \oiii\ (systemic) and \lya\ is at best modestly negative 
and/or may be consistent with zero.

\begin{deluxetable}{c|cccc|c|c}
\tabletypesize{\scriptsize}
\tablecaption{Comparison of \oiii\ and \lya\ redshifts\label{tblz}}
\tablehead{
\colhead{} & \multicolumn{4}{c}{\oiii} & \colhead{\lya} & \colhead{} \\
\colhead{Region} & \colhead{4960.295} & \colhead{5008.24} & \colhead{z$_{avg}$} & \colhead{z$_{corr}$} & \colhead{z} & \colhead{$\Delta$v (\kms)}\\
}
\startdata
C15 (LUCIFER)  &  & 3.0983 &  & \textbf{3.0986 $\pm$ 0.0003494} &  \textbf{3.0976 $\pm$ 0.0004159} & -72.4 $\pm$ 41.6\\
C15 (NIRSPEC) & 3.0973 & 3.0980 & 3.0976 & \textbf{3.0974 $\pm$ 0.0005095}  & \textbf{3.0967 $\pm$ 0.0002070} & -51.3 $\pm$ 42.1\\
C11 (NIRSPEC) & & 3.1002  &  & \textbf{3.0999 $\pm$ 0.0003590} & \textbf{3.1000 $\pm$ 0.0002138} & 5.8 $\pm$ 32.9\\
\enddata
\tablecomments{Comparison of \oiii\ and \lya\ redshifts in C11 and C15 as measured with LUCIFER and NIRSPEC.  z$_{avg}$ is listed when the redshifts from 4960.295 and 5008.24 \AA\ lines were averaged.  z$_{corr}$ is the redshift after correction to the LSR.  $\Delta$v is velocity offset between \oiii\ and \lya.}
\end{deluxetable}

\subsection{\oiii\ Flux in C15 and R2}
 As previously described in Section \ref{oiiifit}, the \oiii\ lines are fit with a symmetric Gaussian plus a constant.  The resulting area under the Gaussian gives us a line flux measurement for each line.  We find an \oiii\ flux of 1.1 $\pm$ 0.1 $\times$ 10$^{-16}$ \lf\ for C15 from LUCIFER (5008 \AA\ line only) .  For C15 in NIRSPEC we find line fluxes of 1.6 $\pm$ 0.5 $\times$ 10$^{-17}$ \lf (4960 \AA) and 4.6 $\pm$ 0.5 $\times$ 10$^{-17}$ \lf (5008 \AA).  Finally, for C11 from NIRSPEC we measure a line flux of 1.6 $\pm$ 0.4 $\times$ 10$^{-17}$ (5008 \AA\ line only).   

Since we do not detect an \oiii\ line in R2 we instead measure an \oiii\ line flux upper limit of $\le$ 3.4 $\times$ 10$^{-17}$ \lf.  LUCIFER has sufficient wavelength coverage to place upper limits on both \oii\ and H$\beta$ in our C15 spectrum.  We derive upper limits for \oii\ and H$\beta$ in C15 of 1.3 $\times$ 10$^{-17}$ and 2.2 $\times$ 10$^{-16}$ \lf\ from our LUCIFER spectrum.  This combination of a measured \oiii\ line flux and upper limits on \oii, and H$\beta$ line flux in C15 allow us to put constraints on the metallicity of the LBG embedded in C15 using  R$_{23}$, where R$_{23}$ is defined as $\frac{\textrm{\oii} + \textrm{\oiii} }{H\beta}$\citep{pag79,kew02}.  For C15, R$_{23} >$ 0.76 given the line flux measurements and limits described above.  NIRSPEC wavelength coverage only allows us to place upper limits on H$\beta$.  We find H$\beta$ upper limits of 3.7 and 7.2 $\times$ 10$^{-18}$ \lf\ for C15 and C11, respectively.

To compute the 3$\sigma$ line flux upper limit for R2 quoted above, we added a mock Gaussian emission line to the spectra to represent \oiii, similar to the procedure in \citet{fink11b}.  The sigma of the Gaussian was fixed to 5.52 \AA, or the $\sigma$ from our faintest \oiii\ detection to date \citep{mc11}.   The area under the mock line was measured using a symmetric Gaussian, this area determines the line flux of the mock line. Then we  determined the noise on the line flux measurement from 1000 Monte Carlo iterations, where the flux array was modified each time by a random amount proportional to the error bars.  We repeated this process, each time  decreasing the area under the mock Gaussian until the signal to noise ratio (SNR) dropped below 5$\sigma$.   The line flux in the mock line where the SNR crossed below  5$\sigma$ determines $\sigma$.  However, because one cannot know, without an nebular emission line measurement for reference, exactly how much, if any the \lya\ line is offset from the \oiii\ line, we repeat this calculation, fixing the mock line at different redshifts to mimic different velocity offsets.  We found the 3$\sigma$ \oiii\ line flux detection limit as an average of this technique from 6 different redshifts corresponding to velocity offsets of 0-500 \kms, in increments of 100 \kms.   A range of 0 - 500 \kms\ was chosen to mirror the magnitude of \lya\ - \oiii\ velocity offsets we have observed of 52 - 342 \kms\ in three z $\sim$ 3.1 LAEs. The range of 3$\sigma$ detection limits over this wavelength range was 3.12 - 3.54  $\times$ 10$^{-17}$  \lf.  The \oii\ and H$\beta$ upper limits described above are found using the same procedure, except the upper limits need only be derived at a single fixed wavelength for each line, the wavelength defined by the \oiii\ redshift.

Given \oiii\ line flux detections in C15 and C11 and an \oiii\ upper limit in R2, we can compare the nebular emission from these subregions.  This is of interest because C15 and C11 are associated with LBGs embedded within the larger LAB1 halo structure, whereas R2, in spite of its stronger \lya\ emission , is not associated with any underlying galaxy \citep{w10}.  The ratio of \lya\ luminosity in C15 to R2 is 0.7 \citep{w10}.  We find that the ratio of \oiii\ in C15 (LUCIFER) to R2 is $\ge$ 3.4 and the ratio of \oiii\ in C15 (NIRSPEC) to R2 is $\ge$ 1.4,  meaning that while R2 is brighter in \lya\, C15 is brighter when looking at  \oiii\ nebular emission.  Finally,  the \oiii\ to \lya\ ratio measured in the region without an LBG is therefore at least 1.9 -- 4.5 times smaller than the same ratio measured in the LBG.  This measurement would likely indicate that something other than star formation is powering the \lya\ emission in region R2 and that there may very well be different sources powering \lya\ emission in different regions of the same blob. 

C11 presents a somewhat different story.  The ratio of \lya\ luminosity in C11 to R2 is 1.0 \citep{w10}.   Then the ratio of \oiii\ in C11 (from NIRSPEC) to R2 is only $\ge$ 0.5.  The \oiii\ to \lya\ in C11 is only 0.06, which is smaller than that same measurement in R2, where the ratio is $\le$ 0.13.  These divergent results are interesting, perhaps giving further credence to the idea that there are different processes at work in different regions of this blob.



\begin{figure}
\begin{tabular}{cc}
\includegraphics[scale=0.3]{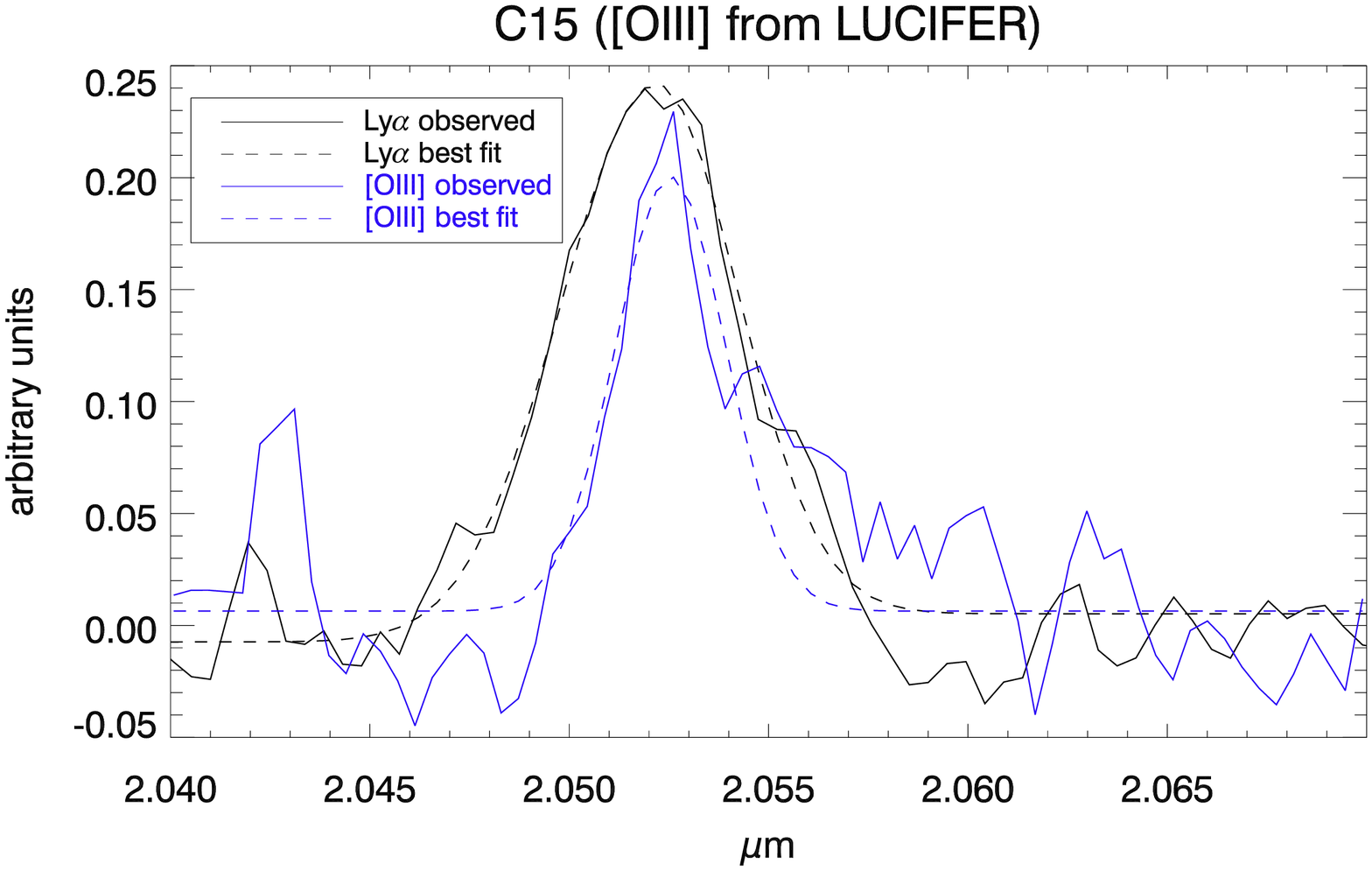} & \includegraphics[scale=0.3]{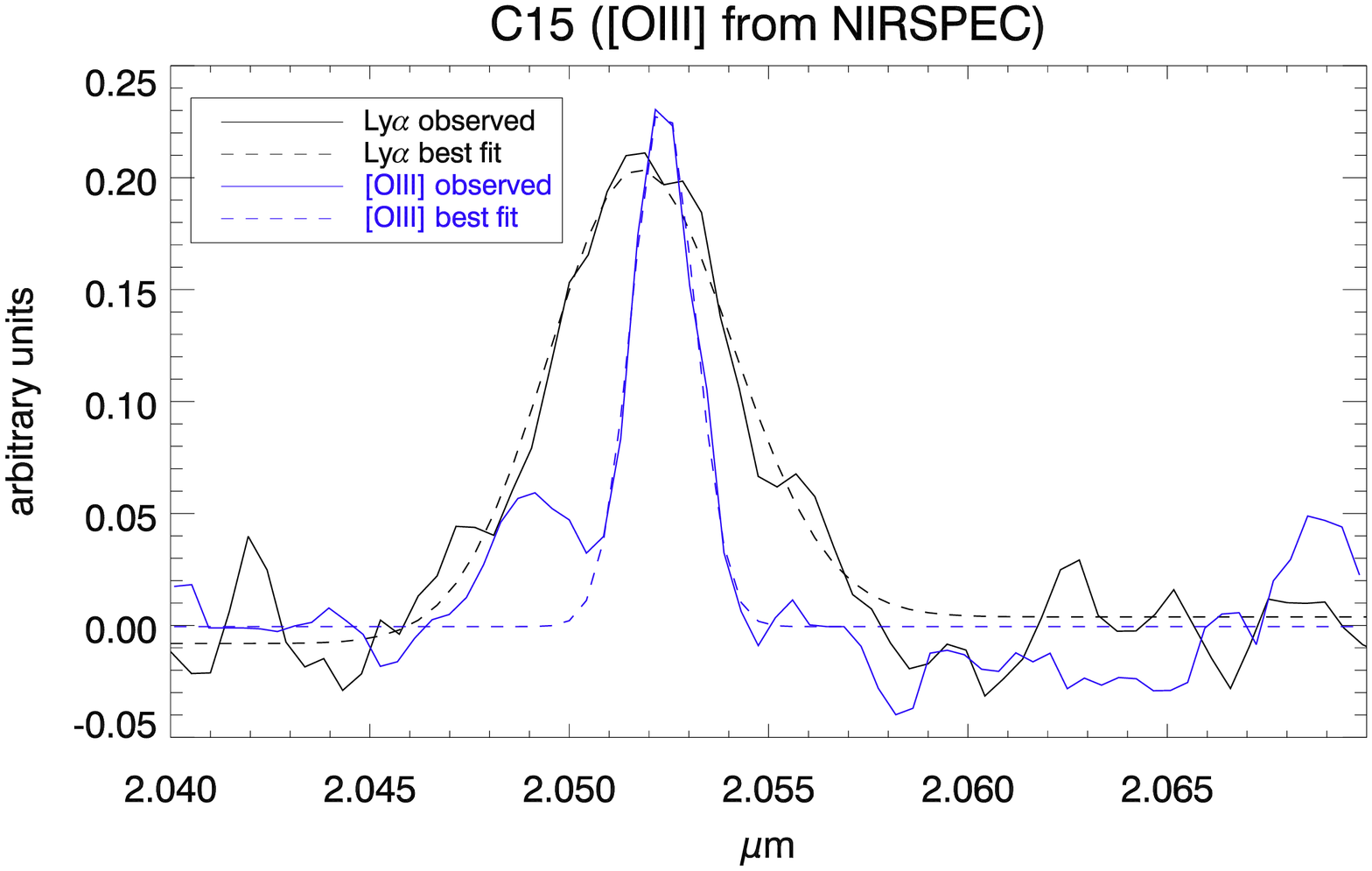}\\
& \includegraphics[scale=0.3]{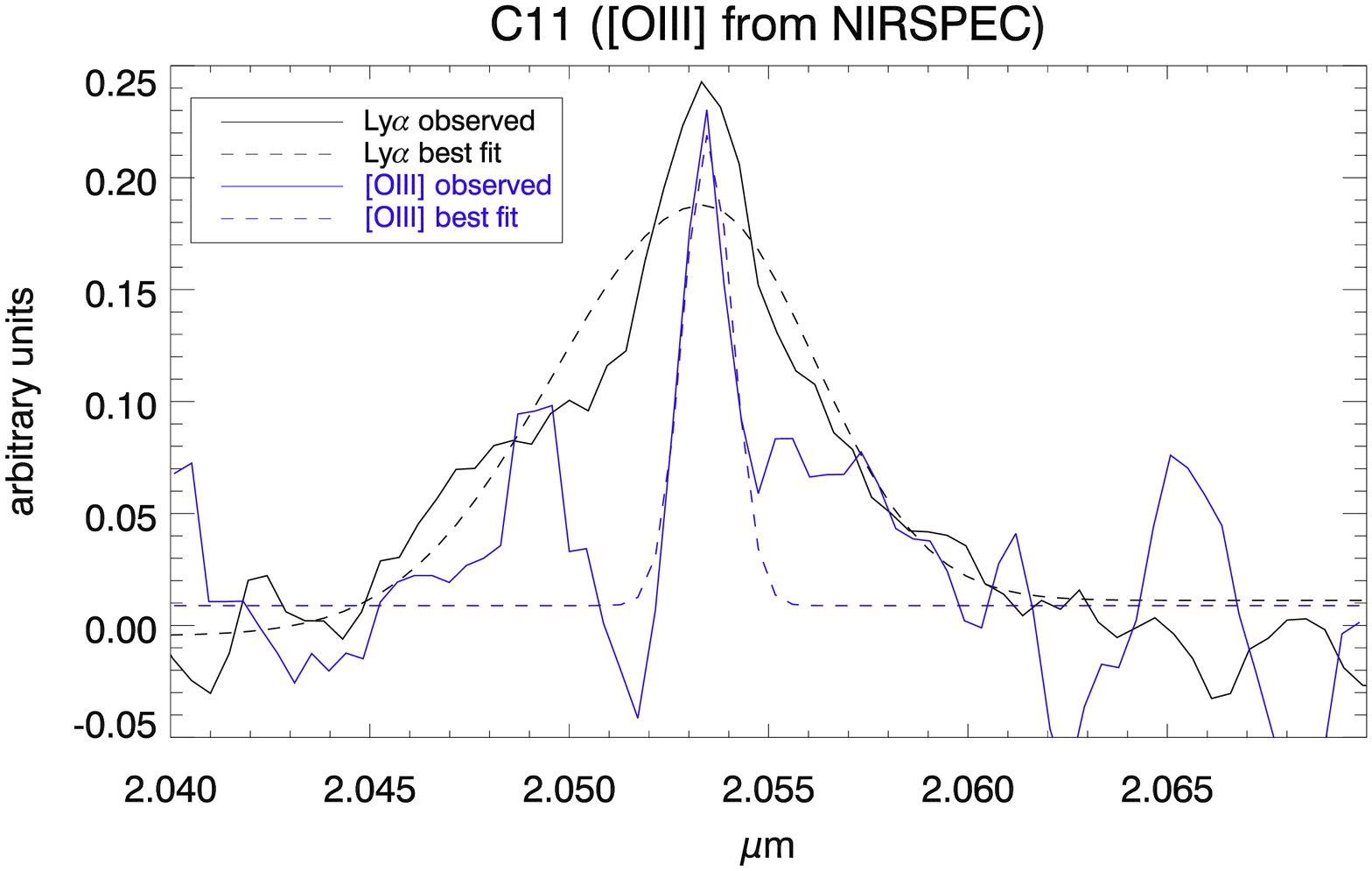} \\
\end{tabular}
\caption{Plot of \lya\ (blue) over \oiii\ (black) where \lya\ has been shifted to \oiii\ frame via \lya$_{\textrm{\oiii}} = \lya_{\textrm{observed}}\times\frac{5008.24}{1215.67}$.  Observed spectra are shown with solid lines, best fit Gaussians for both \lya\ and \oiii\ are shown with dashed lines. Top panel is C15, left plot shows \oiii\ from LUCIFER, right  plot shows \oiii\ from NIRSPEC.  Bottom panel is C11 from NIRSPEC.}\label{voffplots}
\end{figure}

\subsection{Asymmetry and \oiii\ - \lya\ Offset}
We have now measured the velocity offset between \oiii\ and \lya\ in 3 z $\sim$ 3.1 LAEs \citep{mc11} and added 3 $\Delta$v measurements for LAB1 in this work.  Amongst these measurements, the regions C11 and C15 in LAB1 have the smallest absolute $\Delta$v measurements and are the only negative values.  This is in contrast to three z $\sim$ 3.1 LAEs in which we found offsets ranging from 52 -- 342 \kms.  Given this information, we can compare another signature of outflows, namely asymmetry in the \lya\ profile,  with the velocity offset measurements.  

We quantify asymmetry as $\sigma_{red}/\sigma_{blue}$ where $\sigma_{red}$ is the sigma of the red side of the best-fit asymmetric Gaussian and $\sigma_{blue}$ is the sigma on the blue side of the best-fit asymmetric Gaussian, where both are parameters returned by the ARM\_ASYMGAUSSFIT routine.  With this definition, a profile with asymmetry $>$ 1.0 is considered asymmetric, and the asymmetry is dominated by the red-wing.  Objects with asymmetry $=$ 1.0 are symmetric, $\le$ 1.0 have blue-wing dominated asymmetry. The red-wing dominated asymmetry is the expected direction of the asymmetry in the \lya\ line from high-z galaxies, as the red side of the line can be enhanced in the presence of an expanding shell \citep{v06,dij10} and/or by interaction with neutral Hydrogen in the IGM \citep{rho03,daw04}.    Measured in this way C11 has an asymmetry of  0.74 $\pm$ 0.02 and C15 has an asymmetry of  0.9 $\pm$ 0.03 and 1.1 $\pm$ 0.03 (LUCIFER and NIRSPEC apertures, respectively).  An asymmetry of $<$ 1.0, like C11, can suggest infalling material.  However, the \lya-\oiii\ velocity offset of 5.8 $\pm$ 32.9 km/s in C11 is not very supportive of such an interpretation.  In  three z $\sim$ 3.1 LAEs in which we have measured a velocity offset between \oiii\ and \lya\, we find asymmetries of 0.97 $\pm$ 0.1, 1.04 $\pm$ 0.1, and 1.65 $\pm$ 0.1 for LAE7745, LAE27878 and LAE40844, respectively \citep{mc11}.  We add four additional asymmetry data points by using four z $\sim$ 2 LAEs from \citet{hash12}.  \citet{hash12} measured redshifted \lya\ lines in these objects with respect to H$\alpha$ lines in the same objects.  They report velocity offsets of 18 -- 190 \kms\ in their four LAEs, in good agreement with our range of 52 -- 342 \kms\ in LAEs.   We measured the asymmetry of the \lya\ lines presented by \citet{hash12} in the same manner as above, by fitting each line with asymmetric Gaussian and quantifying asymmetry as $\sigma_{red}/\sigma_{blue}$.  Measured in this way the LAEs from \citet{hash12} have asymmetries of 0.53 - 1.7. 

\begin{figure}
\centering
\includegraphics[scale=0.4]{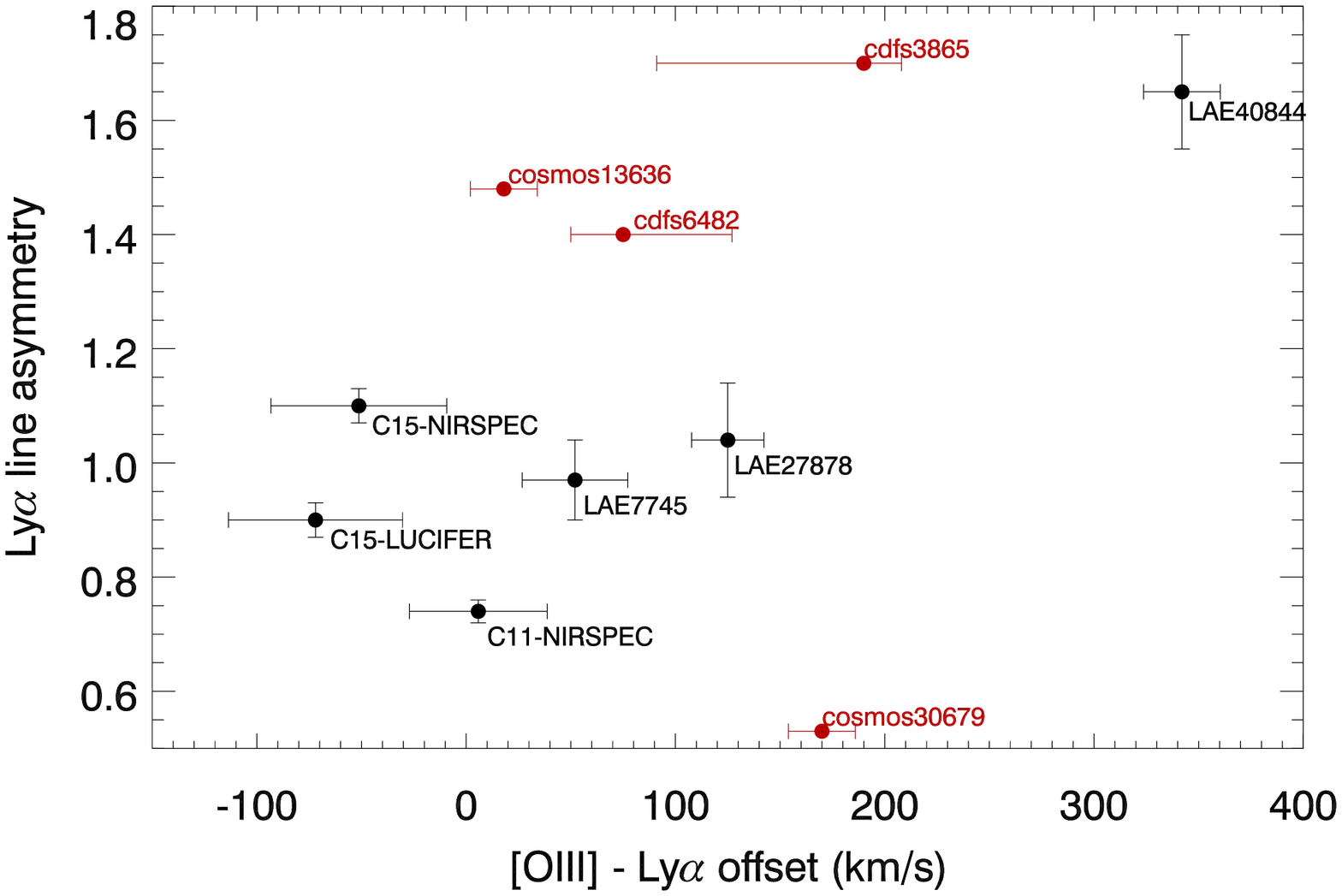}
\caption{Asymmetry of the \lya\ profile as a function of velocity offset in five \lya-emitting objects at z $\sim$ 3.1 (shown in black) and four z $\sim$ 2 objects (shown in red).  Three LAEs from \citet{mc11} are labeled with the prefix `LAE.'  Red points with the prefix `cosmos' or `cdfs' are from \citet{hash12}.  Overall, asymmetry may increase with increasing velocity offset.}\label{fig:vasym}
\end{figure}

Figure \ref{fig:vasym} demonstrates a possible trend where asymmetry in the \lya\ profile increases with increasing velocity offset.  We find a moderate Pearson linear correlation coefficient of 0.427 (P=0.110) between the velocity and asymmetry values, suggesting the trend may be real.  This velocity-asymmetry correlation is not unexpected, as increasing (red-wing dominated) asymmetry in the \lya\ profile is tied to increasing shell expansion velocities in \citet{v06}, when outflows are modeled with a central monochromatic source and an expanding shell.   This is because \lya\ photons are seen as more redshifted by the neutral hydrogen in the shell with higher shell expansion speeds, decreasing the cross-section for interaction \citep{v06}.  A symmetric line, as seen in C15, is not expected in models with large outflows, but could be consistent with static/nearly-static profiles - if  the two symmetric peaks produced from a static slab or shell scenario \citep{v06} are unresolved.  The authors note that the point attributed to LAE40844 strongly influences this correlation.  Removal of this point and recalculation of the Pearson coefficient with the remaining eight points only yields a coefficient of 0.170 (P=0.331), which does not suggest any significant correlation.

\section{DISCUSSION}\label{discussion}
\subsection{Comparison with Blobs at z  $\sim$ 2}
We have measured  $\Delta$v between \oiii\ and \lya\ in two subregions of LAB1 and found that $\Delta$v is modestly negative or consistent with 0 \kms.  These measurements seem to downplay the role of powerful outflows in powering the \lya\ in these regions and could even hint at some infall.  Interestingly, this measurement of $\Delta$v = small is not the first time this phenomenon has been reported in a \lya\ blob, suggesting this is an important phenomenon that must be explored to better understand the nature of high-z LABs.  \citet{ya11} investigated two  z $\sim$ 2.3 LABs, where one blob (called CDFS-LAB02) had $\Delta$v $\sim$ 230 \kms\ and the other blob had $\Delta$v consistent with zero (called CDFS-LAB01).  The velocity offsets reported in \citet{ya11} were measured by comparing the redshift of H$\alpha$ to that of \lya, a very similar tactic to our comparison of \oiii\ to \lya, as both \oiii\ and H$\alpha$ are nebular emission lines and therefore probe the same regions.

\subsection{Comparison to LAEs and LBGs}
Our $\Delta$v measurements, and those of \citet{ya11} , are significantly less than the larger velocity offsets typically seen in Lyman break galaxies (LBGs) and even \lya\ emitting galaxies (LAEs), velocity offsets which are typically interpreted as clear signatures of strong winds in these galaxies. \citet{stei10} report median velocity offsets between H$\alpha$ and \lya\ of 445 \kms\ in 41 z $\sim$ 2.3 LBGs.  And even LAEs, whose typical velocity offsets have been found to be smaller, have $\Delta$v as large as 342 \kms \citep{mc11}.

\citet{stei10} also report a median velocity offset between H$\alpha$ and strong interstellar absorption lines of -164 \kms\ in 86 LBGs, which further supports interpretations of the presence of outflows in LBGs, since the blue-shifted absorption implies absorption in material moving towards the observer.  
\citet{shap06} previously measured a redshift from low-ionization interstellar absorption (LIS) lines in the LBG C11, finding that the absorption lines are offset from the \lya\ line by -380 \kms.  Using the LIS redshift of 3.0962 from \citet{shap06} and comparing this to the \lya\ redshift we derive for C11 yields an offset of $\sim$ 278 \kms.  Comparing our \oiii\ (systemic) redshift for C11 (3.0999) to the LIS redshift, we can estimate that the LIS lines in C11 are offset from \oiii\ by $\sim$ -270 \kms.  This comparison of \lya\ and \oiii\ redshifts to LIS absorption redshifts provides a stronger signature of an outflow than we get when comparing \lya\ and \oiii.  In fact, it is particularly interesting that the magnitude of this second signature of winds (i.e. blueshifted interstellar absorption lines) is so similar when comparing the Steidel LBGs to C11, yet the magnitude of the offset between nebular emission lines (H$\alpha$ or \oiii) and \lya\ offset are so different.  We note that the lack of $\Delta$v between \oiii\ and \lya\ does not have to rule out some outflows in LAB1.  Rather, the lack of a \lya-\oiii\ offset may just imply that outflows are not a significant mechanism for helping \lya\ photons escape.

\subsection{Previous Studies of LAB1}
This phenomena, i.e. $\Delta$v = small and/or = 0 \kms, leads to the question of whether the lack of substantial velocity offset between H$\alpha$-\lya\ or \oiii-\lya\ in these blobs in fact rules out outflows or if there is some, yet to be understood phenomena, that damps or erases this particular wind signature.  This question is particularly relevant given the recent report from \citet{hay11} that there is polarized radiation emanating from LAB1, polarization that is indicative of scattering of \lya\ photons at large radii with respect to their site of production.  This may be a sign of outflows helping to drive the \lya\ photons to these large radii, but we are not seeing the velocity offsets between nebular emission lines and \lya\ that we  would expect to see if this was the case, velocity offsets we have been able to see in other objects at similar redshifts believed to have strong winds.  In addition, \citet{bow04} and \citet{w10} both measure a velocity shear in the \lya\ emission from C11 and C15.  As the authors point out, such a velocity shear could be consistent with infalling gas, outflowing gas and/or rotation of the system, and such scenarios cannot be differentiated from the \lya\ data alone.  While both papers use this velocity shear to argue in favor of the presence of outflows in C11 and C15, we can report no signature of such outflows when we compare the redshifts of \oiii\ and \lya, a comparison that has proven to be a useful diagnostic of winds in LAEs and LBGs at similar redshifts.

\subsection{Comparison to Radiative Transfer Models}
\begin{figure}[!hb]
\epsscale{0.85}
\plotone{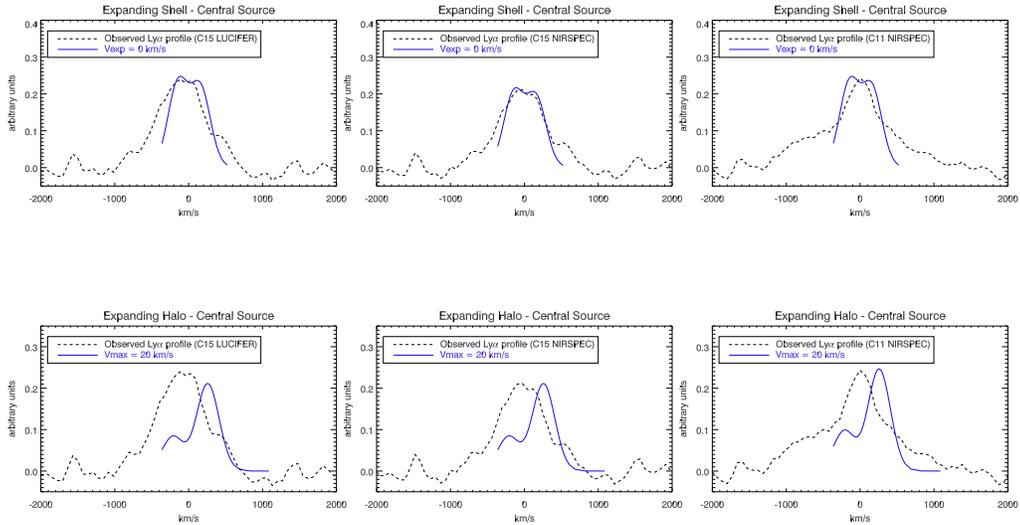}
\caption{\lya\ profiles (black dashed line) compared to V06 outflowing shell and halo models (solid blue). V06 models have been convolved with a Gaussian to match instrumental resolution of SAURON. Top row shows comparison to model in V06 Figure 14. Bottom row shows comparison to model in V06 Figure 7.   While these models and those in Figure \ref{figv06_2} represent the closest matches to the observed profiles, clearly none are an excellent fit.}\label{figv06}
\end{figure}

\begin{figure}[!hb]
\epsscale{0.85}
\plotone{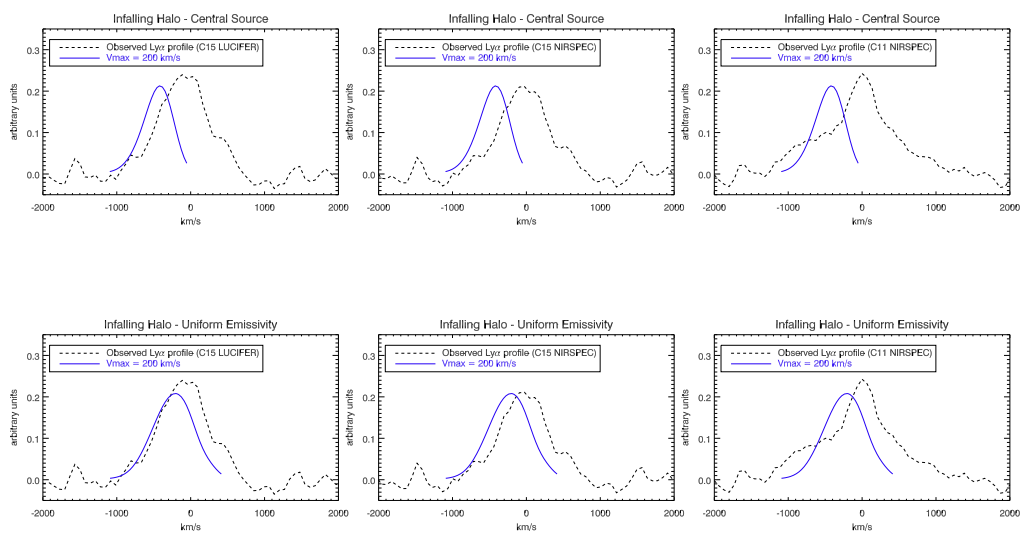}
\caption{\lya\ profiles (black dashed line) compared to V06 infalling halo models (solid blue). V06 models have been convolved with a Gaussian to match instrumental resolution of SAURON. Top row shows comparison to model in left panel of Figure 5 in V06, bottom row shows comparison to model in right panel of Figure 5.  While these models and those in Figure \ref{figv06} represent the closest matches to the observed profiles, clearly none are an excellent fit.}\label{figv06_2}
\end{figure}

We explored available \lya\ radiative transfer models to see if there were any models that might shed light on the physical conditions that could lead to a modestly negative $\Delta$v or $\Delta$v of $\sim$ 0 \kms\ between \oiii\ (or H$\alpha$) and \lya.  We focused on the \lya\ profiles produced by \citet{v06} (henceforth V06) from their 3D Monte Carlo \lya\ radiative transfer code.  V06 explore a variety of physical conditions and geometric orientations to explore the diversity of \lya\ profiles that arise from different conditions.  See V06 for extensive details on these models and the model parameters.  We find, however, that none of the models presented in V06 are in great agreement with our observations (or the observations of \citet{ya11}).  The only models that are marginally consistent with our observations are those that have two significant \lya\ peaks, where the centroid of those two peaks is centered at v $\sim$ 0, with one peak redward of v= 0, and the other blueward.  Convolved with an appropriate Gaussian to mimic the instrumental resolution of SAURON (290 \kms), these models somewhat resemble the \lya\ profiles presented here.  This double-peaked, centered at v $\sim$ 0 profile occurs when a central monochromatic source is embedded in a static slab (with or without dust, V06 Figure 3), a central monochromatic source sits in a non-expanding shell (V06 Figure 14), and when a single monochromatic source sits in an expanding shell with very small velocity gradient (V$\max$ = 20 \kms, V06 Figure 7).  The closest match (see Figure \ref{figv06}) is from V06, Figure 14 with a static shell.  In addition, we explored whether the infalling halo models of V06 might agree with our observations, these models are presented in Figure \ref{figv06_2}.  There may be some qualitative agreement between the models and observed profiles here, but the models explored are too far blueward of v=0 to match the \lya\ profiles we present here.  In either scenario (expanding or infalling) the main peak is expected to be asymmetric, which is  inconsistent with the  \lya\ profiles of at least C15 presented here.  
We contend that more modeling of \lya\ radiative transfer with direct applications to the observations we have presented here and those observations presented in \cite{ya11} needs to be done, to better understand the physical conditions, geometry, and kinematics that can produce single peaked \lya\ lines, with $\Delta$v values that are modestly negative or consistent with zero.

\section{CONCLUSION}
We have measured \oiii\ in two subregions of LAB1, C11 and C15, regions that are associated with underlying LBGs within the larger halo structure.  We have quantified the velocity offset between \oiii\ and \lya\ redshifts in these regions, finding that both measurements are modestly negative or consistent with zero.  This is an intriguing result since powerful outflows have been proposed as possible ways to explain the luminosity and large spatial extent of LABs.  We cannot completely rule out the presence of strong winds and outflows in LAB1, but we can state that we do not see two typical markers of their presence.  (1) The aforementioned result that we do not find that \lya\ is redshifted with respect to \oiii\  in the two LAB1 subregions C15 and C11.  And (2) we do not measure strong red-wing dominated asymmetries in the \lya\ profiles of these objects, in contrast with  z $\sim$ 2 and z $\sim$ 3.1 LAEs where the asymmetry of the \lya\ line appears to increase with increasing velocity offset between \oiii\ (or H$\alpha$) and \lya.  If outflows are present in LAB1, they do not appear to be a crucial mechanism driving \lya\ escape.

In addition to the conclusion above, we have placed an upper limit on \oiii\ line flux from region R2, a subregion of LAB1 unassociated with any known galaxy and compared this to the \oiii\ flux from subregion C15, which is associated with an LBG.  We find that in spite of the strong \lya\ emission present in R2, the \oiii\ flux from C15 is stronger than that of R2 by at least 1.4 -- 2.5 times.  This measurement may indicate that diverse sources of \lya\ emission may be responsible for powering different regions even within the same blob.  

\acknowledgements
The LBT is an international collaboration among institutions in the United States, Italy and Germany. LBT Corporation partners are: The University of Arizona on behalf of the Arizona university system; Instituto Nazionale di Astrofisica, Italy; LBT Beteiligungsgesellschaft, Germany, representing the Max-Planck Society, the Astrophysical Institute Potsdam, and Heidelberg University; The Ohio State University, and The Research Corporation, on behalf of The University of Notre Dame, University of Minnesota and University of Virginia.  

This work was supported by a NASA Keck PI Data Award (award number is 1408709), administered by the NASA Exoplanet Science Institute. Data presented herein were obtained at the W. M. Keck Observatory from telescope time allocated to the National Aeronautics and Space Administration through the agency's scientific partnership with the California Institute of Technology and the University of California. The Observatory was made possible by the generous financial support of the W. M. Keck Foundation.  The authors wish to recognize and acknowledge the very significant cultural role and reverence that the summit of Mauna Kea has always had within the indigenous Hawaiian community. We are most fortunate to have the opportunity to conduct observations from this mountain.

NSF grant AST-0808165 also provided  funding for this work.  

Thanks to Mark Richardson for various helpful conversations.  Thanks also to Richard Bower for his contributions to the SAURON data and for sharing this data.

\end{document}